\documentclass[prd,aps,twocolumn,floatfix]{revtex4}
\usepackage{amssymb,graphicx}
\usepackage{epsfig}
\usepackage{amsmath}
\usepackage[usenames]{color}

\begin{document}

\title{Dynamical evolution of fermion-boson stars}

\author{
Susana Valdez-Alvarado$^{1}$,
Carlos Palenzuela$^{2}$,
Daniela Alic$^{3}$,
L. Arturo Ure\~na-L\'opez$^{1}$}
\affiliation{
$^{1}$Departamento de F\'isica, DCI, Campus Le\'on, CP 37150,
    Universidad de Guanajuato, Le\'on, Guanajuato, M\'exico, \\
$^{2}$Canadian Institute for Theoretical Astrophysics, Toronto, Ontario M5S 3H8,
 Canada, \\
$^{3}$Max-Planck-Institut f\"ur Gravitationsphysik,
    Albert-Einstein-Institut, Potsdam-Golm, Germany}

\begin{abstract}
Compact objects, like neutron stars and white dwarfs, may accrete
dark matter, and then be sensitive probes of its presence. These
compact stars with a dark matter component can be modeled by a perfect
fluid minimally coupled to a complex scalar field (representing a
bosonic dark matter component), resulting in objects known as
fermion-boson stars. We have performed the dynamical evolution of
these stars in order to analyze their stability, and to study their
spectrum of normal modes, which may reveal the amount of dark matter
in the system. Their stability analysis shows a structure similar to
that of an isolated (fermion or boson) star, with equilibrium
configurations either laying on the stable or on the unstable
branch. The analysis of the spectrum of normal modes indicates the
presence of new oscillation modes in the fermionic part of the star,
which result from the coupling to the bosonic component through
the gravity.
\end{abstract}

\maketitle

\section{Introduction \label{sec:introduction-}}

Scalar fields are of great interest in several fields of physics. In
high energy physics they arise naturally in several unification
theories, such as scalar-tensor theories of gravitation
from string theory. In cosmology, they have been considered to provide
inflationary solutions in the early universe and an alternative
explanation for dark energy. In addition, they have also been proposed 
as strong candidates of dark matter, the matter that is
responsible for the formation and evolution of structures in the
Universe. For the latter type of models, one of the possibilities
assumes that dark matter is composed by bosonic particles which may
condensate into macroscopic self-gravitating objects (i.e.,
self-gravitating Bose-Einstein condensates) commonly known as
\emph{boson stars}. Since the seminal paper in the late sixties by
Ruffini and Bonazzola~\cite{Ruffini:1969qy}, boson stars in General
Relativity have been extensively studied in may different contexts
(for a recent review see~\cite{lrr-2012-6}).

On the other hand, the formation of either a boson or a fermion star
would be susceptible to subsequent mixture by fermions/bosons,
and this fact opens a whole new possibility for the formation of
objects made of both fermionic and bosonic particles. Even if one of
these objects is formed in a medium absent of either bosonic or
fermionic particles, the latter may be accreted in later stages. In
particular, bosonic dark matter particles may accrete on compact
stars, and depending on the model considered, their effects on the
star will be different and possibly observable. 

In the context of WIMPs, if the dark matter is self-annihilating,
the released energy due to the WIMP annihilation inside the neutron star can
increase the temperature and be observable in old stars
\cite{Kouvaris:2007ay}. If it does not self-annihilate, the dark
matter will cluster in a small region at the center of the neutron
star, increasing their compactness and ultimately leading to a
gravitational collapse \cite{Bertone:2007ae}. Neutron stars may be
therefore sensitive indirect probes of the presence of dark matter,
and can be used to set constraints both on the density and on the
physical properties of dark matter.

Recent studies investigate possible changes in the structure of the star
in the presence of dark matter, by using a two-fluid
model~\cite{Leung:2011zz}.  In this paper, we perform a similar
analysis by modeling systems which contain a fermionic compact star (we consider
it to be a neutron star) and a bosonic dark matter component
represented by a boson star. The resulting objects are known
as fermion-boson stars. These mixed stars were first introduced by
Henriques et. al.~\cite{Henriques:1989ar} (and further studied
in~\cite{Lopes:1992aw,Henriques:2003yr}), where the fermionic matter
was described by a perfect fluid with the Chandrasekhar equation of
state, while the bosonic component is modeled by using a real quantized
scalar field as in~\cite{Breit:1983nr}.
The bosons and the fermion particles are coupled only through gravity
(notice however that non-minimal couplings with the scalar field
can arise in other scenarios, such as in neutron stars
with hidden extra dimensions~\cite{1990CQGra...7.1009L} or in
tensor-scalar theories of gravitation~\cite{1993PhRvL..70.2220D}).
We will perform a dynamical analysis of these mixed stars by using
a simple polytropic equation of state, as it is  standard for cold
neutron stars, and a complex scalar field to describe the bosonic
component. 

The equilibrium configurations for either an isolated boson or fermion 
stars are described, respectively, by the central value of the scalar 
field $\phi_c$, and the central value of the fluid density 
$\rho_c$~\cite{Ruffini:1969qy,Shapiro:1983du}. These configuration are
therefore characterized by a single parameter $\sigma_c$, so in this
case there are stability theorems~\cite{Thorne:1965,Straumann:1984}
which indicate that the critical mass (separating the unstable from
the stable branch) is located at the extrema $\partial M/\partial \sigma_c = 0$.
However, the mixed fermion-boson stars are parameterized not by one, but
by two quantities $(\phi_c,\rho_c)$. This implies that  the analysis of stability
is more complicated than in the isolated star case, since the previous stability
theorems can not be directly applied. One can still analyze their stability,
among other alternatives, by studying the radial perturbations of these equilibrium
configurations and then analyzing the eigenvalues of these modes in
the linearized equations as
in~\cite{Gleiser:1988rq,Gelmini:1988sf,Gleiser:1988ih,Jetzer:1988vr,Lee:1988av},  
or by evolving dynamically these perturbations by solving the full 
non-linear equation of 
motion~\cite{Kusmartsev:2008py,Balakrishna:1998pa,Jetzer:1997zx,Bernal:2009zy}. In~\cite{Henriques:1990xg,Henriques:1989ez}   
Henriques et al. described a method to perform the analysis of 
stability of the boson-fermion stars by using the binding energy and
the number of bosonic and fermionic particles as a function of the two
free parameters. In this paper, we propose a similar criterion,
and our results are compared with the full numerical solution of the
equations of motion. In addition to the stability analysis, we 
follow the migration of a star from an unstable to the stable branch,
a process observed already in isolated boson and fermion stars.
Finally, we study the dependence of the quasi-normal modes of the
mixed star with respect to their fraction of bosonic matter. 

The paper is organized as follows. In Sec.~\ref{sec:formalism} we
introduce the formalism used to obtain the set of evolution equations
that describes the spacetime geometry and the boson-fermion matter
contents. In Sec.~\ref{sec:id} we describe how to construct the
initial data for the boson-fermion stars, and propose a method to find
the stability of theses objects. The results of the dynamical
evolution for equilibrium configurations (i.e., both stable and
unstable) are presented in
Sec.~\ref{sec:numerical_simulations}, together with the spectrum of
the quasi-normal modes of the stable stars. Finally, conclusions and
final remarks are presented in Sec.~\ref{sec:concluding_remarks}.
Throughout this paper we use that the indices are $a,b,.$ taken
to run from 0 to 3, while indices $i,j,..$ run from 1 to 3.
We also adopt the standard convention for the summation over repeated
indices.

\section{Formalism}
\label{sec:formalism}

Fermions minimally coupled to bosons can be modeled by considering a
stress-energy tensor  with contributions from a perfect fluid and a
complex scalar field, in the form 
\begin{eqnarray}\label{stressenergy} 
  T_{ab} &=& T_{ab}^{\rm (fluid)} + T_{ab}^{\rm (\phi)} \, , \\
  T_{ab}^{\rm (fluid)} &=& \left[ \rho_o\left(1+\epsilon\right) + P
  \right] u_a u_b + P g_{ab}  \, , \\ 
  T_{ab}^{\rm (\phi)} &=&
  \frac{1}{2}\left[ \partial_{a}\phi^\ast \partial_{b} \phi
    + \partial_{a} \phi \partial_{b} \phi^\ast \right] - \nonumber \\ 
  & &\frac{1}{2} g_{ab} \left[ \partial^{\alpha}
    \phi^\ast \partial_{\alpha} \phi + m^2 |\phi|^2 \right] \, .  
\end{eqnarray}
The perfect fluid is represented by the fermionic physical (primitive)
variables, namely the pressure $P$, rest-mass density $\rho_o$, 
internal energy $\epsilon$, and four-velocity $u^a$, whereas the complex
scalar field $\phi$ describes a Bose-Einstein condensate of bosonic
particles of mass $m$. The fluid and the scalar field do not interact
directly, and are only coupled through gravity, as it is expected
for WIMPS. The equations of motion for the fluid and the scalar field
are obtained from {\em the conservation laws} of the stress-energy
tensor and the baryonic number
\begin{equation} 
\nabla_a T^{ab}_{\rm (fluid)} = 0 \, , \quad \nabla_a (\rho_o u^a) = 0
\, ,  
\label{HD4d}
\end{equation} 
and the Klein-Gordon equation 
\begin{equation} 
 \nabla_a \nabla^a \phi = m^2 \phi \,, 
\label{KG4d}
\end{equation} 
which together with the Einstein equations $G_{ab} = 8\pi T_{ab}$
constitute the system of equations governing the dynamics. 

We restrict our study to spherically symmetric stars, and then
consider the time-dependent metric
\begin{equation}
  ds^2 = -\alpha^2(t,r)\, dt^2 + g_{rr}(t,r)
  dr^2+r^2g_{\theta\theta}(t,r)\, d\Omega^2.
\label{metricageneral}
\end{equation}
The evolution equations for the spacetime are obtained by considering
the $Z3$ formulation of the Einstein equations~\cite{Alic:2007ev}, which 
introduces the following independent quantities to form a first order
system of equations,
\begin{eqnarray}
A_r &=& \frac{\partial_r\alpha}{\alpha}\, , \,\,\,
D_{rr}{}^{r}=\frac{g^{rr}}{2}\partial_rg_{rr}\, ,\,\,\,
D_{r\theta}{}^{\theta}=\frac{g^{\theta\theta}}{2}\partial_rg_{\theta\theta}\,,
\nonumber \\
K_{r}{}^{r} &=& -\frac{1}{2 \alpha} \frac{\partial_{t}g_{rr}}{g_{rr}} \, , \,\,\,
K_{\theta}{}^{\theta} = -\frac{1}{2 \alpha}
     \frac{\partial_{t}g_{\theta\theta}}{g_{\theta\theta}}\, .
\end{eqnarray}
The full system of equations for this formulation is included in
appendix~\ref{app:z3}. The remaining freedom in the choice of coordinates
of the line element (\ref{metricageneral}) is related to the prescription
for the lapse function, and a common option is the harmonic slicing condition
\begin{equation}
\partial_t\alpha=-\alpha^2trK \,,
\label{eq:lapse}
\end{equation}
where $trK=K_r{}^r+2K_{\theta}{}^{\theta}$. By using the metric
(\ref{metricageneral}), the equations of motion for the perfect fluid
(\ref{HD4d}) and the scalar field (\ref{KG4d}) can be written
explicitly as:
\begin{widetext}
\begin{subequations}
\label{eq:ekghd}
\begin{eqnarray}
\partial_t(\sqrt{\gamma}D) &=& -\partial_r(\sqrt{\gamma}\alpha
v^{r}D) - \frac{2}{r}\sqrt{\gamma}\alpha v^rD,
\\
\partial_t(\sqrt{\gamma}U) &=& -\partial_r(\sqrt{\gamma}\alpha
\tilde{S}^r)+\sqrt{\gamma}\alpha\Big[\tilde{S}_r{}^rK_r{}^r+2\tilde{S}_{\theta}{}^{\theta}K_{\theta}{}^{\theta}-
\tilde{S}^r\Big(\frac{2}{r}+A_r\Big)\Big],
\\
\partial_t(\sqrt{\gamma}\tilde{S}_r) &=& -\partial_r(\sqrt{\gamma}\alpha
\tilde{S}_r{}^r)+\sqrt{\gamma}\alpha\Big[\tilde{S}_r{}^r\Big(D_{rr}{}^r-\frac{2}{r}\Big)+
2\tilde{S}_{\theta}{}^{\theta}\Big(\frac{1}{r}+D_{r\theta}^{\theta}\Big)-UA_r\Big], 
\\
\partial_t\phi_t &=& \partial_r(\alpha\sqrt{g^{rr}}\phi_r)+\alpha\sqrt{g^{rr}}\Big[2\Big(D_{r\theta}{}^{\theta}+
\frac{1}{r}\Big)\phi_r+2\sqrt{g_{rr}}K_{\theta}{}^{\theta}\phi_t-m^2g_{rr}\phi\Big],
\end{eqnarray}
\end{subequations}
\end{widetext}
where $\sqrt{\gamma}=\sqrt{g_{rr}}g_{\theta\theta}$ and we have introduced
the auxiliary fields
\begin{eqnarray}
\phi_r &=& \partial_r\phi\, , \,\,\,
\phi_t=\frac{\sqrt{g_{rr}}}{\alpha}\partial_t\phi\,,
\end{eqnarray}
to reduce the Klein-Gordon equation to first order in space and time.

The evolution of the fluid is described in terms of the {\it conserved}
variables, namely the mass density $D$, the momentum density
$\tilde{S}_{r}$ and the energy density $U$. They are related to the
primitive variables (i.e., the rest-mass density $\rho_o$, the pressure
$P$ and the velocity $v_{r}$) by the following relations
\begin{eqnarray}\label{eq:con2prim}
   D &=& \rho_{0} W\,, \quad  U = h W^{2} - P \,, \quad 
   \tilde{S}_{r} = h W^2 v_{r}\,,
\end{eqnarray}
where $h=\rho_{0}(1+\epsilon)+P$ is the enthalpy and 
$W=1/\sqrt{1-v^{r}v_{r}}$ the Lorentz factor. In the right-hand-side
of their evolution equations, the spatial projections of the
stress-energy tensor take the form,
\begin{eqnarray}\label{eq:rhs_conserved}
   \tilde{S}_{r}{}^{r} &=& h W^2 v_{r} v^{r} + P \,, \quad 
   \tilde{S}_{\theta}{}^{\theta} = P \,.\nonumber
\end{eqnarray}
During the evolution, the relations (\ref{eq:con2prim}) must be
inverted in order to obtain the primitive physical quantities
(which are necessary for computing the rhs) from the conserved
evolved fields. In general, this conversion can not be performed
analytically, so appendix~\ref{app:conservadas} explains in detail
our numerical algorithm for obtaining the primitive fields.
 
\section{Initial data}
\label{sec:id}

Initial data for the fermion-boson stars involves the intrinsic
metric $g_{ij}$ and extrinsic  curvature $K_{ij}$ on a given hyper-surface,
as well as the fermionic fluid configuration in terms of
its primitive variables $(\rho,\epsilon,v^i)$ and the bosonic scalar
field $\phi$. Assuming a static spherically symmetric metric in
Schwarzschild  coordinates 
\begin{equation}
  ds^2 = - \alpha^2(r) dt^2 + a^2(r) dr^2 + r^2 d\Omega^2 \, ,
\end{equation}
a harmonic form of the scalar field $\phi(t,r)=\phi(r)e^{-i\omega t}$
\footnote{Notice that, since regular and time-independent 
localized scalar field solutions are not stable in three spatial
dimensions~\cite{Derrick:1964ww,1966JMP.....7.2066R}, an harmonic
complex scalar field is the simplest way to obtain a compact and
stationary star.},
and a star in hydrostatic equilibrium with $v_{r}=0$, the following
system of ODEs is obtained:
\begin{widetext}
\begin{subequations}
\begin{eqnarray}
\frac{da}{dr} &=& \frac{a}{2} \left\{ \frac{1}{r}(1-a^{2})+4\pi G
  r\left[\left(\frac{\omega^{2}}{\alpha^{2}}+m^{2}\right)a^{2}\phi^{2}(r)+
 \Phi^{2}(r)+2a^{2} \rho(1+\epsilon) \right] \right\} \, ,\label{equi1}\\
\frac{d\alpha}{dr} &=& \frac{\alpha}{2}\left\{\frac{1}{r}(a^{2}-1)+4\pi G
  r\left[\left(\frac{\omega^{2}}{\alpha^{2}}-m^{2}\right)a^{2}\phi^{2}(r)+
  \Phi^{2}(r)+2a^{2}P\right]\right\}\, ,\label{equi2}\\
\frac{d\phi}{dr} &=& \Phi(r)\, ,\label{equi3}\\
\frac{d\Phi}{dr} &=& \left(m^{2}-\frac{\omega^{2}}{\alpha^{2}}\right)
a^{2}\phi -\left[1+a^{2}-4\pi Ga^{2}r^{2}\left(m^{2}\phi^{2}+
\rho(1+\epsilon)-P\right)\right] \frac{\Phi}{r}\, ,\label{equi4}\\
\frac{dP}{dr} &=& -\left[\rho(1+\epsilon)+P\right]
                  \frac{\alpha^\prime}{\alpha}\,. \label{equi5}
\end{eqnarray}
\end{subequations}
\end{widetext}
The system is completed by choosing the equation of state (EoS) that
relates the pressure with the other fluid quantities. As it is
standard in simple models of cold stars, we will adopt here a
polytropic equation of state $P=K \rho^{\Gamma}$, with the particular
choice of $\Gamma=2$ and $K=100$, which corresponds to masses and
compactness in the range of neutron stars~\cite{Shapiro:1983du}.

We will use units such that $c=1$, and the variables can be renormalized
to absorb the factors $G$ and $m$, so that the basic scale of the stars
will be given by $\{K,\Gamma\}$. The final system is an eigenvalue problem
for the frequency of the boson star $\omega$ as a function of two parameters;
the central value of the scalar field $\phi_{c}$ and the density
of the fluid $\rho_{c}$. This system can be solved by using the Shooting
Method~\cite{Press:1992zz}. 

The appropriate boundary conditions for the scalar field and metric
functions are obtained by imposing the conditions of regularity at the origin
and asymptotic flatness at infinity. The condition at $r=0$ 
for the fluid pressure is obtained from the polytropic EoS
as a function of $\rho_{c}$. Thus, the full boundary
conditions are
\begin{subequations}
\begin{eqnarray}
a(0) &=& 1,\label{funmet} ~~~~~
\alpha (0) = 1,\label{lapse} ~~~~~
\phi(0) = \phi_{c},~~\label{phicentral}\\
\Phi(0) &=& 0,\label{primphi} ~~~~~
P(0) = K\rho_{c}^{\Gamma},~~\label{press}\\
\lim_{r\rightarrow\infty}\alpha(r) &=& \lim_{r\rightarrow\infty}\frac{1}{a(r)},
~~\label{finalpha}\\
\lim_{r\rightarrow\infty}\phi(r)&\approx&\ 0,~~~~~\label{finphi}
\lim_{r\rightarrow\infty}P(r) = 0.~\label{fipres}
\end{eqnarray}
\end{subequations}

After the solution is found, a change of coordinates from
Schwarzschild to maximal isotropic ones is performed
\begin{equation}
ds^2=-\alpha^2(\tilde{r})d\tilde{t}^2
    + \psi^4(\tilde{r}) \left( d\tilde{r}^2+\tilde{r}^2d\Omega^2
    \right) \,,
\end{equation}
which are more convenient for our numerical evolution and for future
comparisons in three dimensions. All of our simulations will be shown
in these coordinates, and for simplicity we will substitute $\tilde{r}
\rightarrow r$ hereafter.

The total gravitational mass is computed by the asymptotic value
of the metric coefficients
\begin{equation}
 M_T=\lim_{r\rightarrow\infty}\frac{r}{2}\left(1 -
\frac{1}{\alpha^{2}}\right) \,.
\label{masa}
\end{equation}
The $U(1)$ symmetry in the Lagrangian of the scalar field
ensures the conservation of a Noether charge which can be associated
with the number of bosons $N_B$~\cite{Bernal:2009zy,Henriques:1990xg}.
Correspondingly, the conservation of baryonic number allows to define
a number of fermions $N_F$. These quantities can be computed
by integrating their densities, 
\begin{eqnarray}
\frac{\partial N_{B}}{\partial r} =
\frac{4\pi a\omega\phi^{2}r^{2}}{\alpha} \,, ~~~
\frac{\partial N_{F}}{\partial r} = 4\pi a\rho r^{2} \,.
\label{radioferbos}  
\end{eqnarray}
Therefore, the radius of the fermionic/bosonic parts of the star can
be defined as the surface containing $99\%$ of the corresponding
particles.

\subsection{Boson  and fermion stars \label{sec:boson-fermion-stars}}

As a basic test of our initial data implementation we 
compare our equilibrium configurations with previously published
results for isolated boson stars and fermion stars, which are the 
limits of our system of equations when $\rho_c \rightarrow 0$ and
$\phi_c \rightarrow 0$ respectively.
 
\begin{figure}[ht]
  \includegraphics[height=.24\textheight]{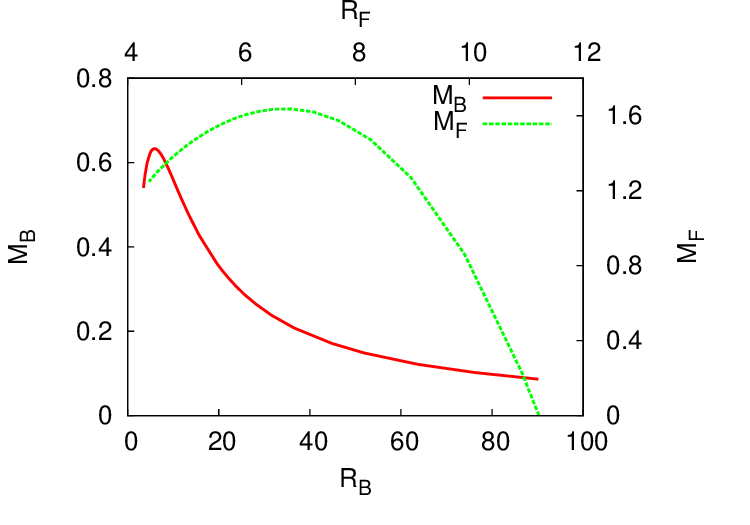}
  \caption{{\it Initial data of isolated stars}. The total masses
    of the boson $M_B$ and fermion $M_F$ stars, as functions of their
    corresponding radius $R_{99}$. The maximum mass agrees in each
    case with previous results found in the literature, namely
    $M_{max}=0.633$ for boson stars, and $M_{max}=1.637$ for fermion
    stars.}
  \label{fig:inicial}
\end{figure}

Fig.~(\ref{fig:inicial}) shows the total mass $M_T$ of boson stars and
fermion stars as a function of the corresponding radius $R_{99}$. In
agreement with the results of previous works, we have found that the
maximum mass $M_{max}$ (i.e., the value of the mass that separates the
stable $M_T < M_{max}$ from the unstable $M_T > M_{max}$
configurations) in the case of boson stars is $M_{max}=0.633$, whereas
for fermionic stars is $M_{max}=1.637$ with $\Gamma=2$ and $K=100$. 
 
\subsection{Mixed boson-fermion stars \label{sec:mixted-boson-fermion}}

\begin{figure}[ht]
 \includegraphics[height=.24\textheight]{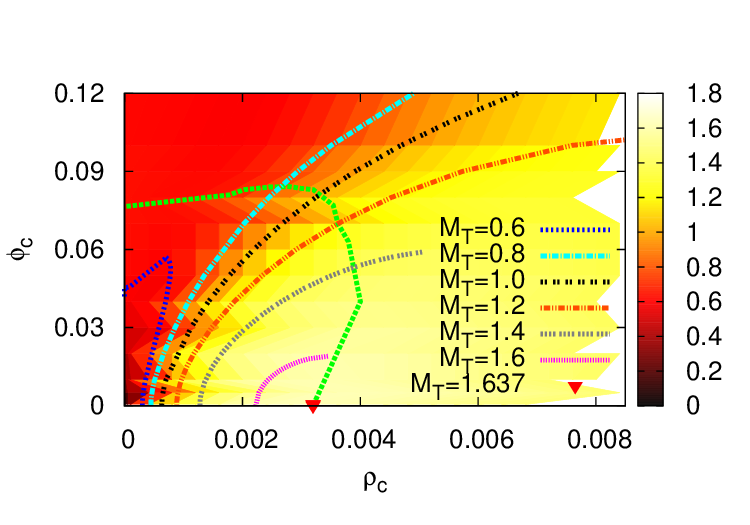}
   \caption{{\it Initial data of mixed fermion-boson stars}.
     The total mass of the equilibrium configurations of the mixed
     stars, $M_T$, as a function of $\phi_c$ and $\rho_c$. 
     The maximum mass, for a given
     value of $\rho_c$, is always found when $\phi_{c}=0$, implying 
     that $M_{max}=1.637$ is the maximum mass value for any boson-fermion
     star in this study.
     The (green) solid line that intersects
     the axes is the stability boundary discussed in the text, see
     also Fig.~\ref{fig:migracion}.}
  \label{fig:masas}
\end{figure}

As we mentioned before, the equilibrium configurations of mixed
boson-fermion stars are more involved and depend on the two parameters
$\phi_{c}$ and $\rho_{c}$. The total mass of the stars as a function
of these parameters is plotted in Fig.~\ref{fig:masas}, showing that
the maximum mass is obtained for the isolated neutron star case
(i.e.,when $\phi_{c}=0$). This is a direct consequence of our choice
of the parameters $\{K,\Gamma\}$ in the equation of state, that sets the
scale and the compactness of the mixed stars. With the current choice,
the stars will be composed predominantly by fermions, which can produce
stars with much higher compactness than boson stars.

\begin{figure}[ht]
  \includegraphics[height=.24\textheight]{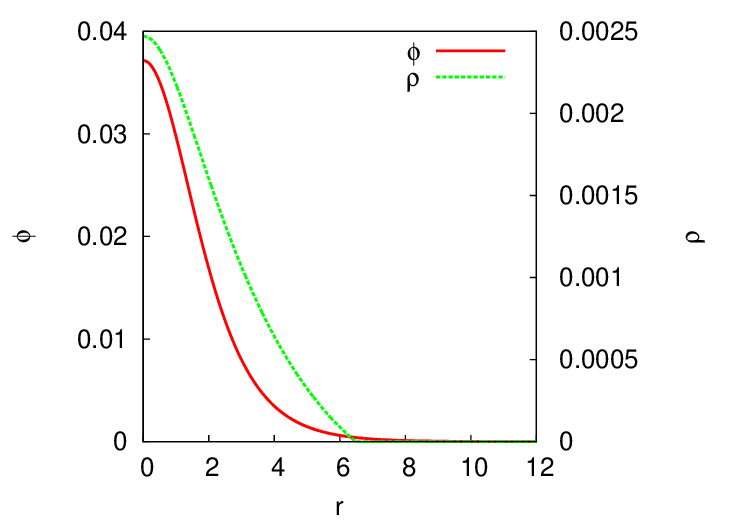}
  \includegraphics[height=.24\textheight]{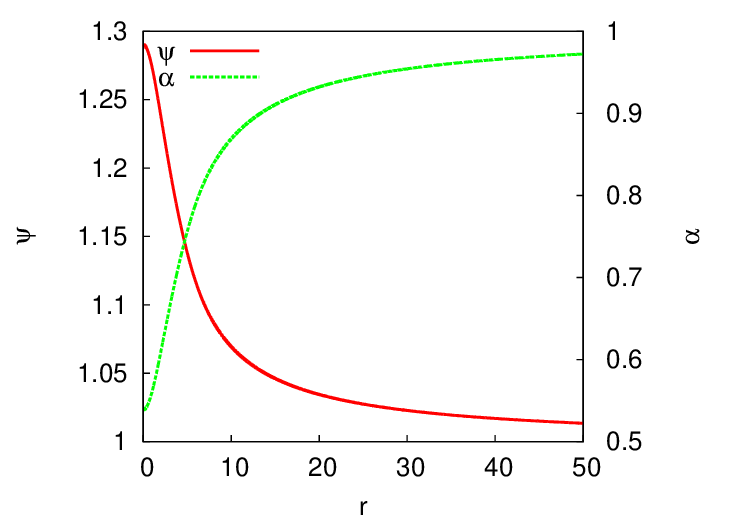}
  \caption{{\it Initial data of mixed fermion-boson stars.}
    The profiles of the scalar field $\phi(r)$, the fermionic
    density $\rho(r)$, and the conformal factor $\psi(r)$
    for one typical configuration corresponding to $N_B/N_F=0.1$
    and $M_T=1.4$.}
  \label{fig:perfiles}
\end{figure}

The profiles of the different non-trivial fields for a representative
case are plotted in Fig.~\ref{fig:perfiles} , which clearly satisfy the
regularity conditions at the origin and asymptotic flatness. The presence
of the fermionic fluid produces a deeper gravitational potential than
the one produced solely by the boson star, therefore contracting the bosonic
component to a smaller radius, comparable to the one of the fermionic matter.

\begin{figure}[htbp!]
  \includegraphics[height=.23\textheight]{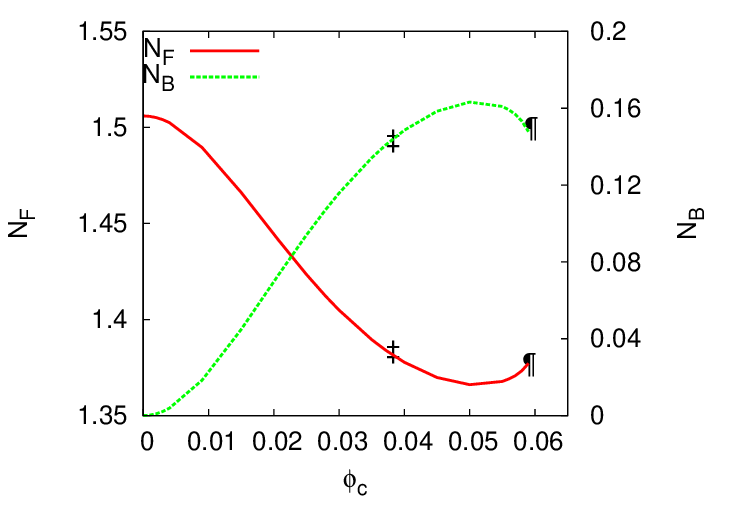} \\
  \includegraphics[height=.23\textheight]{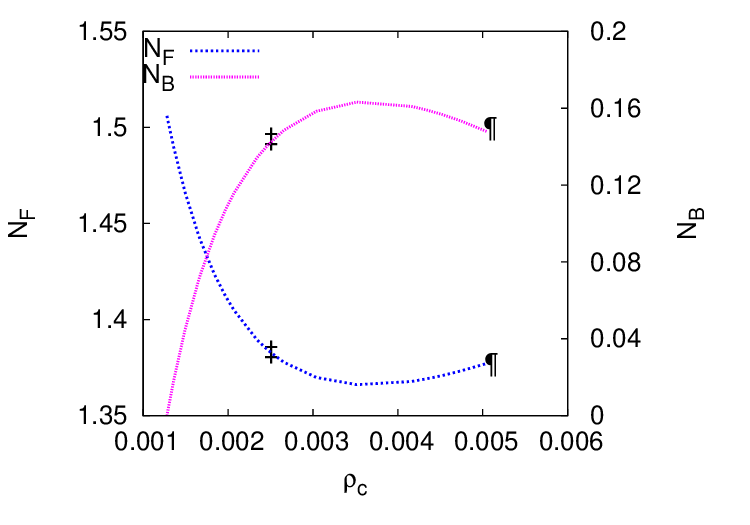} \\
\includegraphics[height=5.6cm,width=6.0cm,angle=0]{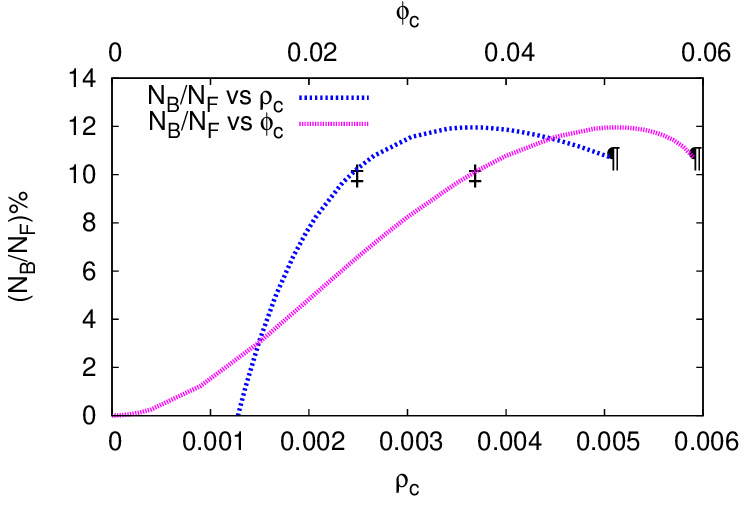} 
\caption{{\it Initial data of mixed fermion-boson stars}. The number
  of fermions $N_F$ and bosons $N_B$ for the equilibrium
  configurations as a function of $\phi_{c}$ (top panel) and
  $\rho_{c}$ (middle panel) corresponding to the fixed total mass
  $M_T=1.4$. The position of the  maximum/minimum corresponds to the
  critical point which separates the stable and the unstable
  solutions. The two configurations considered in the next section are
  marked, one on each side of the maximum/minimum, corresponding to
  $N_B=10\%N_F$($\ddag$) and $N_B = 10.7\% N_F$
  ($\mathparagraph$). (Bottom panel) The boson-to-fermion
  ratio, $N_B/N_F$ for $M_T=1.4$; the critical configuration
  corresponds to the maximum value $N_B/N_F=12\%$.}
  \label{fig:particulas}
\end{figure}

For a fixed value of the total mass $M_T$, we find that the number of
bosons $N_B$ increases for $\phi_c \ge 0$,
reaches a maximum, and then decreases. Notice that since the total
mass is kept fixed, the central density $\rho_c$ must change as we
vary $\phi_c$. The number of fermions $N_F$ has consequently the
complementary behavior to that of $N_B$: it decreases until reaching a minimum and then
increases. The same profiles are observed in these quantities
when they are represented as a function of $\rho_c$ instead of $\phi_c$.
This behavior is illustrated
in Fig.~\ref{fig:particulas}, where the number of particles is plotted
as a function of $\phi_c$ and $\rho_c$ for the configurations with
mass $M_T=1.4$. 

Following usual nomenclature, we call 'critical point'
an equilibrium fermion-boson solution -- described by
$\{M,N_F,N_B\}$-- which separates the stable from the unstable
configurations. This transition is signaled by the passing through
zero of the lowest eigenvalue in the (radial) perturbations of the
fermion-boson star.

The stability analysis for the boson-fermion stars, which can be
performed for instance by solving the perturbed equations of motion,
is much more complicated than for isolated boson or fermion stars
(see for instance~\cite{Henriques:1989ez,Henriques:1990xg,Lopes:1992aw,
Henriques:2003yr}). The main reason, as it was mentioned before, is that
these mixed configurations have two free parameters (i.e., the central
values of the scalar field $\phi_c$ and the density of the perfect fluid
$\rho_c$) instead of just one, so that the stability theorems for a
single parameter solutions can not be directly applied.

Instead of performing again this stability analysis, we
propose a different way to find critical configurations. Our
criterion is based on the studies made in~\cite{Henriques:1990xg},
in which the authors realized that in a critical point (i.e., where
the eigenvalue of the perturbation vanishes) there must be a
direction $\mathbf{n}$ such that the directional derivatives of
$\{M,N_F,N_B\}$ vanish 
\begin{equation}
  \label{eq:directional}
  \left. \frac{dM}{d\mathbf{n}} \right|_b =
  \left. \frac{dN_B}{d\mathbf{n}} \right|_b =
  \left. \frac{dN_F}{d\mathbf{n}} \right|_b = 0 \, ,
\end{equation}
where the subscript $b$ means the value of the quantities at the critical
point. The direction $\mathbf{n}$ at the stability boundary is
tangential to the level curves of constant $M$, $N_B$, and $N_F$;
formally speaking, the direction $\mathbf{n}$ is orthogonal to the
gradient of the functions at the boundary, $\mathbf{n} \perp
\left. \nabla (M_T,N_B,N_F) \right|_b$. Therefore, the stability
boundary can be found by drawing contours onto the plane
$(\phi_c,\rho_c)$ for fixed values of the particle numbers, and
looking for the points where these curves meet and are tangential one
to each other; this was method used in~\cite{Henriques:1990xg}.

In addition, we have noticed --as also did the authors
in~\cite{Henriques:1990xg}-- that the curves of constant mass $M_T$
osculates the curves of constant particle numbers, so that the
stability boundary can be found by surveying the behavior of $N_B$ and
$N_F$ while keeping fixed the value of $M_T$. More precisely, a level
curve of constant total mass, $M_T(\rho_c,\phi_c) = M_0$, implicitly
defines the trajectory $\phi_c = \phi_c(\rho_c,M_0)$, and then the
derivatives of the particle numbers along this given trajectory are
\begin{equation}
  \label{eq:derivatives}
  \frac{dN_B}{d\rho_c} = \left[ \nabla N_B \cdot \mathbf{s} \right]
  (\rho_c) \, ,
  \quad \frac{dN_F}{d\rho_c} = \left[ \nabla N_F \cdot
    \mathbf{s}\right] (\rho_c) \, ,
  \nonumber 
\end{equation}
where $\mathbf{s}(\rho_c) = (1,d\phi_c/d\rho_c)$ is the velocity
vector of the level curve $\phi_c = \phi_c(\rho_c,M_0)$ at any given
point. As we approach the boundary line $\mathbf{s} \to \mathbf{n}$,
the derivatives in Eqs.~(\ref{eq:derivatives}) must vanish at
the critical point as stated by
Eqs.~(\ref{eq:directional}). Consequently, the equilibrium critical
configurations manifest themselves at the extreme values of the number
of particles when surveyed along a level curve of constant total mass.

\begin{figure}[htbp!]
  \includegraphics[height=.24\textheight]{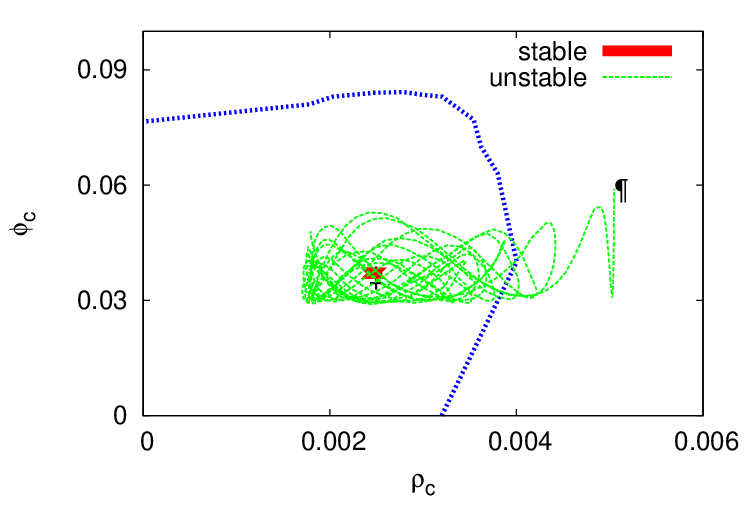}
  \caption{{\it Initial data of mixed fermion-boson stars}.
    Regions of stability/instability for the equilibrium
    configurations of the mixed boson-fermion stars, according to the
    criterion of maximum/minimum of the number of bosons/fermions for
    a fixed $M_T$, see also Fig.~\ref{fig:particulas} and the text
    for more details. Notice that the values at the axes coincide with
    the fermion and boson expected critical values, which are
    $\rho_c=3.2\times10^{-3}$, and $\phi_c=7.65\times10^{-2}$,
    respectively. We mark the two configurations corresponding to
    $N_B=10\%N_F$($\ddag$) and $N_B = 10.7\% N_F$ ($\mathparagraph$),
    whose stability is studied numerically in Figs.~\ref{fig:estable}
    and~\ref{fig:inestable}, respectively. The first one is stable and
  remains in the same state (point), whereas the second is unstable and
  migrates towards the stable region.}
  \label{fig:migracion}
\end{figure}

For the particular case $M_T=1.4$ displayed in
Fig.~\ref{fig:particulas}, the critical configuration is obtained when
$N_B = 0.163$ and $N_F=1.37$. We can see that the critical point also
corresponds to a maximum of the boson-fermion ratio $N_B/N_F$, whose
critical value for $M_T=1.4$ is $N_B/N_F =12\%$. We applied this
recipe for values in the range $0.633 \leq M_T \leq 1.637$, where the
limiting values are established by the critical boson and fermion
cases, respectively, and the solution space of stable/unstable
configurations using this criterion is shown in
Fig.~\ref{fig:migracion}. As noted before in ~\cite{Henriques:1990xg},
stable configurations lie inside this boundary line, where the known stable cases of boson and fermion stars are
found.

Notice that the procedure described above is quite general, and we
could have used the level curves of any of the functions involved. For
instance, if we would have taken the level curve for $N_B(\rho_c,\phi_c)= N_0$,
we could have surveyed $M_T,N_F$ and found that the critical
configuration corresponds to their extreme values. Moreover, either
$\rho_c$ or $\phi_c$ could have been used as the independent
variable, as it is shown in Fig.~\ref{fig:particulas}.

Summarizing, our criterion is based on the analytical work
in ~\cite{Henriques:1990xg}, but takes advantage of the fact that
critical configurations appear as critical points of the particle
numbers when the total mass is held fixed. The configurations with the
number of bosons (fermions) on the left of the maximum (minimum) are
stable configurations, while configurations that are on the right of
the maximum (minimum) are unstable. These results are validated
through numerical simulations presented in the next section, where we
describe the evolution of the two equilibrium configurations marked
with symbols in Fig.~\ref{fig:particulas}, one with $N_B=10\%N_F$,
which is on the left of the critical ratio, and another with
$N_B=10.7\%N_F$, which is on the right of the critical ratio. Our
simulations show that the first configuration is stable (i.e., small
perturbations are bounded, and the system remains in the same state
throughout the entire evolution), whereas the second one is unstable
and the system changes to a different configuration.

This behavior is also shown in Fig.~\ref{fig:migracion}
where we can see that the stable configuration stays in the
stable region, while the unstable configuration leaves the unstable
region and migrates towards the stable one. This suggests that the
maximum/minimum values of the aforementioned curves may mark the
existence of a critical configuration for a given fixed mass.

\section{Numerical Simulations}
\label{sec:numerical_simulations}

In this section we analyze the dynamics of mixed stars, and address
different issues like the stability of these systems or their spectrum
of normal modes.  In order to determine the properties of the mixed
star equilibrium configurations described in the previous section, we
performed long-term numerical evolutions of the discretized
Einstein-Klein-Gordon-Hydrodynamic system~(\ref{eq:ekghd}).

We write the system in flux conservative form
\begin{equation}
     \partial_t {\bf U} + \partial_k F^k ({\bf U}) = S({\bf U}) \, ,
\end{equation}
so that we can apply numerical algorithms based on Finite Volume methods. 
The spatial discretization of the geometry and the boson fields is
performed using a third order accurate Finite Volume
method~\cite{Alic:2007ev}, which can be viewed as a fourth order
finite difference scheme plus third order adaptive dissipation. The
dissipation coefficient is given by the maximum propagation speed in
each grid point. For the fluid matter fields, we use a High Resolution
Shock Capturing method with Monotonic-Centered limiter. The time
evolution is performed through the method of lines using a third order accurate
Strong Stability Preserving Runge-Kutta integration scheme
\cite{ShuOsher88}, with a Courant factor of $\Delta t/\Delta r = 0.25$
so that the Courant-Friedrichs-Levy~(CFL) condition dictated by the
principal part of the equations is satisfied. Most of the simulations
presented in this work have been done with a spatial resolution of
$\Delta r = 0.01$, in a domain with outer boundary situated at $r =
600$. We use maximally dissipative boundary conditions for the
spacetime variables and the boson fields, and outflow boundaries for the
fluid matter fields.

\subsection{Stable boson-fermion
  stars \label{sec:stable-boson-fermion}}

\begin{figure}[ht]
  \includegraphics[height=.24\textheight]{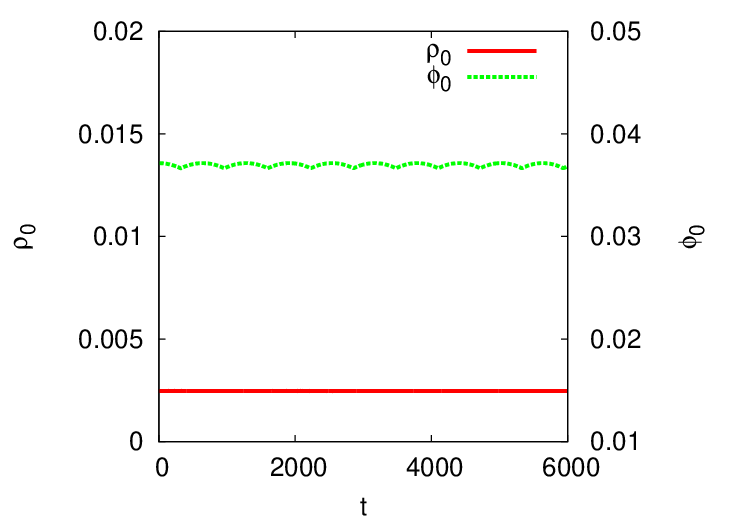}
  \includegraphics[height=.24\textheight]{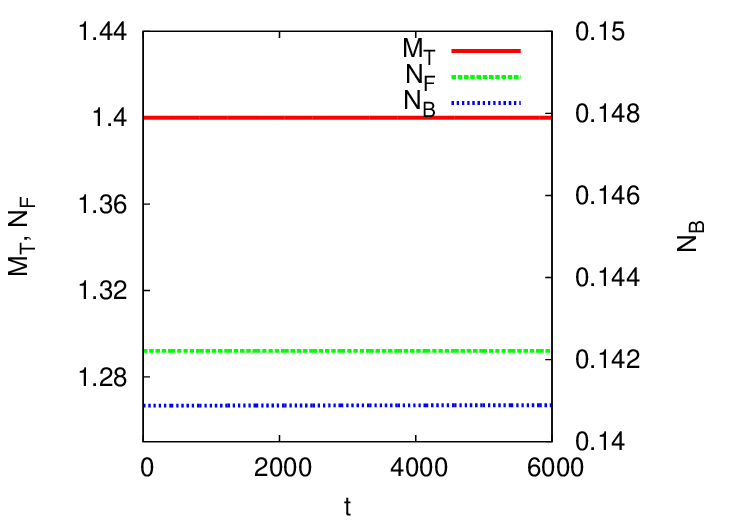}
  \caption{{\it Evolution of stable fermion-boson stars}
   The central values of the density and the (peaks of the oscillatory)
    scalar field
   (top), and the mass and the number of bosonic and fermionic particle
   (bottom). All the quantities remain very close to their initial values,
   suggesting that the star is stable against perturbations.}
  \label{fig:estable}
\end{figure}

The dynamical evolution of the mixed equilibrium
solution corresponding to $N_B=10\% N_F$ and total mass $M_T=1$.$4$,
is shown in Fig.~\ref{fig:particulas}. Since it is located on the
left of the critical values, we expect this configuration to be
stable.

The evolution displays a combination of the behaviors that are typical
for isolated boson and fermion stars. The scalar field oscillates with
its characteristic eigenfrequency, while the fluid density oscillates
slightly around its initial state due to the perturbation introduced
by the numerical truncation errors. The values of the peaks of the 
oscillatory scalar field $\phi^{max}_0$ and the fluid density $\rho_0$
at the center of the star are plotted as a function of time in the top panel of
Fig.~\ref{fig:estable}, while the total mass $M_T$ and the number of
particles $N_B$, $N_F$ are displayed in the bottom panel.

\begin{figure}[ht]
  \includegraphics[height=.24\textheight]{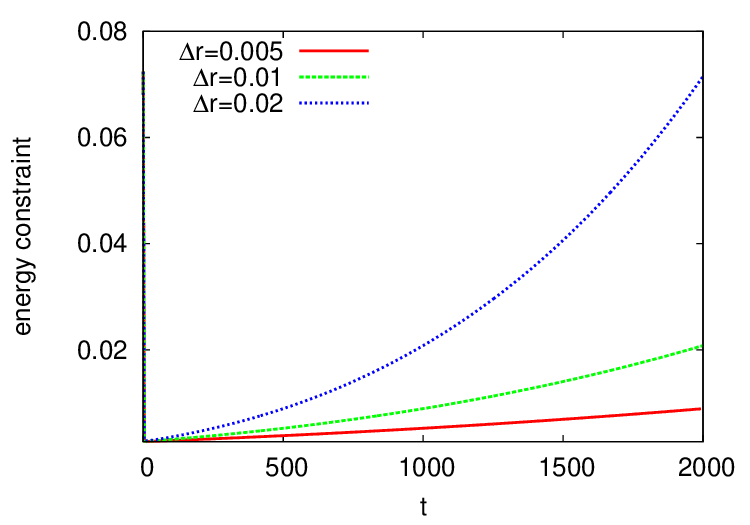}
   \caption{The energy constraint~(\ref{eq::ham}) for three different
     resolutions $\Delta r = 0$.$005$ (red), $\Delta r = 0$.$01$
     (green), and $\Delta r  = 0$.$02$ (blue), showing second order
     convergence.}
  \label{fig:convergence}
\end{figure}

These quantities remain very close to their initial value for many
dynamical times (except for a tiny drift due to numerical
dissipation), indicating that the configuration is indeed stable. In
order to asses the robustness and accuracy of our numerical
implementation, we have evolved this configuration with three
different spatial resolutions $\Delta r = (0.02, 0.01, 0.005)$, in a
domain of $r = 600$, for
$t \approx 2000$, finding
that the numerical solution converges at second order. The energy
constraint~(\ref{eq::ham}) is small during the evolution and
converges to zero, as it is
shown in 
Fig.~\ref{fig:convergence}. 

\subsection{Unstable boson-fermion stars \label{sec:unstable-stars-}}

\begin{figure}[htbp!]
  \includegraphics[height=.25\textheight]{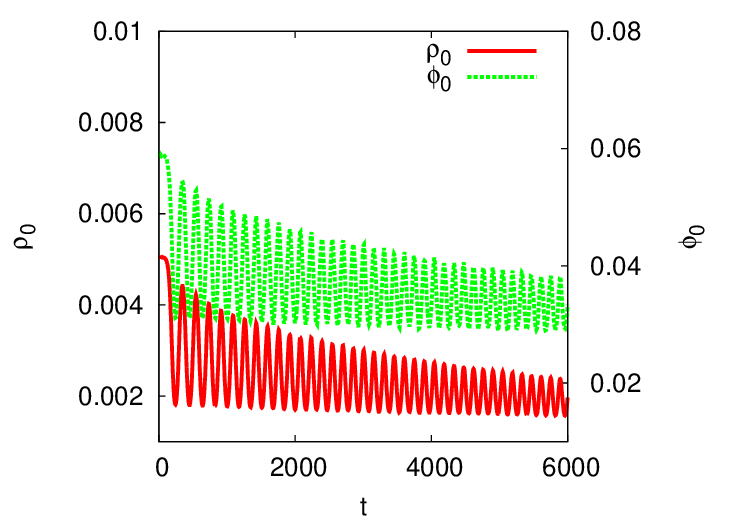}
  \includegraphics[height=.25\textheight]{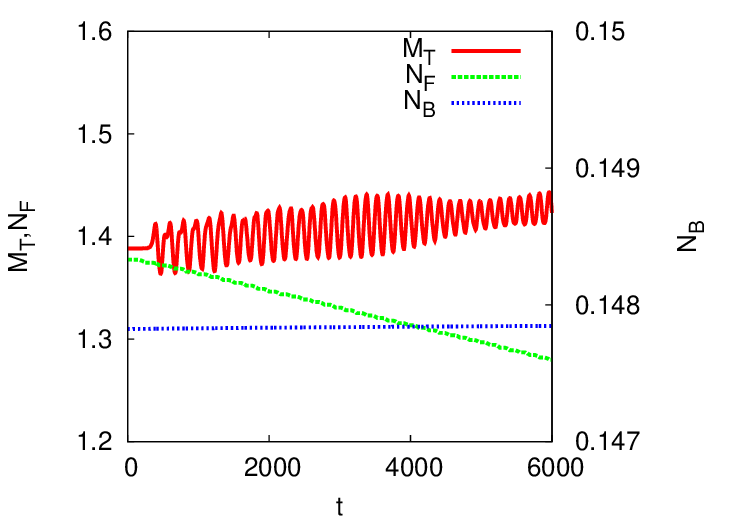}
  \caption{{\it Evolution of unstable fermion-boson stars}
   Same as fig.~\ref{fig:estable} for a star on the right of the critical
   curve. The central values of the density and (the peaks of) the scalar field 
   depart quickly from their initial values, indicating that the 
   star is unstable. The evolution becomes non-linear and describes the
   migration of the star from the unstable to the stable branch.}
  \label{fig:inestable}
\end{figure}

The numerical evolution of the equilibrium configuration with
$N_B=10$.$7\% N_F$ and $M_T=1$.$4$ presents a more dynamical behavior.
This configuration lies on the right of the critical values of the
number of particles $N_F$ and $N_B$ in Fig.~\ref{fig:particulas}, indicating
that it is unstable under perturbations. 

The initial stage of the evolution is similar to the previous case of
a stable star, with the scalar field oscillating mainly with its
eigenfrequency, and the neutron star oscillating due to the
perturbation introduced by the inherent numerical truncation
errors. However, these oscillations grow rapidly in amplitude, driving
the dynamics to a non-linear regime; the star is eventually migrating
from the unstable to the stable branch. The central values $\rho_0$
and the maximum $\phi^{max}_0$, the total mass $M_T$, and the number
of particles $N_B$ and $N_F$, are plotted in
Fig.~\ref{fig:inestable}. The central values show large variations
which are damped slowly, and finally will settle down
onto a new stable configuration with practically the same number of
bosonic and fermionic particles.

Notice however that the large oscillations in the central density
induces significant variations with the same frequency in the radius of
the star. The interaction of the expanding
and contracting star's surface with the atmosphere (i.e., the low
density fluid populating the star's exterior) produces an artificial
small loss of the baryonic mass during the migration that can be observed
in Fig.~\ref{fig:inestable}.

\subsection{Quasi-Normal Modes of the stable
  stars \label{sec:quasi-normal-modes}}

As it has already been mentioned, the fermion-boson star will
oscillate around its stable configuration due to the
perturbations introduced by the numerical truncation errors, in
a similar way as an isolated fermion or boson star. These
perturbations will excite the characteristic modes of the mixed star,
so that the oscillations will be a superposition of normal modes, each
one with a characteristic frequency.

By analyzing the central oscillations of the different fields, and 
in particular, of the central density of the star, we can study 
the structure of the normal modes of the fermion-boson stars.
The frequencies of the normal modes are well-known for both isolated
neutron and boson stars, but they have not been yet studied for
mixed stars. The pulsations of compact objects are of great importance
for relativistic astrophysics, because they offer the possibility of
extracting information about the star (for instance the radius, mass and 
equation of state) from the detection of the associated gravitational
waves (see~\cite{lrr-2003-3} for a review). Although our spherical
symmetry assumption only allow us to study radial modes (i.e., the
fundamental mode and its overtones), it is still representative to
show how these modes may change in a neutron star in the presence  of
a bosonic dark matter component that couples to fermions only through
gravity.

\begin{table*}
 \begin{tabular}{|l|cccc|cccc|}
\hline
\hline
Branch & $ N_B/N_F(\%)$ & ~~$\phi_c$ & ~~$\rho_c$ & ~~$\omega_B$ & ~~$R_T$ & ~~$N_B$ & ~~$R_B$ & ~~$R_F$ \\
\hline
stable & $0.0$  & ~~$0.0$ & ~~$1.27\times10^{-3}$ & ~~$0.0$   & ~~$9.10$ & ~~$0.0$    & ~~$0.0$  & ~~$8.55$ \\
stable & $2.5$  & ~~$1.33\times10^{-2}$ & ~~$1.45\times10^{-3}$ & ~~$0.736$ & ~~$8.76$ & ~~$0.037$  & ~~$6.36$& ~~$8.23$ \\
stable & $5.0$  & ~~$2.06\times10^{-2}$ & ~~$1.67\times10^{-3}$ & ~~$0.718$ & ~~$8.39$ & ~~$0.073$  & ~~$5.96$& ~~$7.88$ \\
stable & $7.5$  & ~~$2.80\times10^{-2}$ & ~~$1.97\times10^{-3}$ & ~~$0.694$ & ~~$7.99$ & ~~$0.107$  & ~~$5.52$& ~~$7.50$ \\
stable & $10.0$ & ~~$3.63\times10^{-2}$ & ~~$2.42\times10^{-3}$ & ~~$0.661$ & ~~$7.49$ & ~~$0.141$  & ~~$5.02$& ~~$7.08$ \\
unstable & $10.7$& ~~$5.90\times10^{-2}$ & ~~$5.05\times10^{-3}$ & ~~$0.533$ & ~~$6.22$ & ~~$0.147$ & ~~$3.60$& ~~$5.82$ \\
\hline
\hline
\end{tabular}
\caption{Properties of the fermion-boson star models used in the
  simulations. All the stars have a total mass $M_T=1.4$. The columns
  report: the fraction of boson particles, the central value of the
  scalar field $\phi_c$, the central density $\rho_c$, the internal
  frequency of the scalar field $\omega_B$, the total radius of the
  star $R_T$, the number of bosonic particles $N_B$, and the radius of
  the bosonic and the fermionic components, $R_B$ and $R_F$,
  respectively. Notice that the largest fraction of $N_B/N_F$ for a
  stable configuration is reached for the maximum value of  $N_B$ and
  the minimum value of $N_F$, but the precise value of the fraction
  depends on the value of the total mass (in the present case we get
  $N_B/N_F\approx 12\%$); larger ratios $N_B/N_F$ can be obtained for
  smaller values of $M_T$.} 
\label{tab:stars}
\end{table*}

\begin{figure}[htbp!]
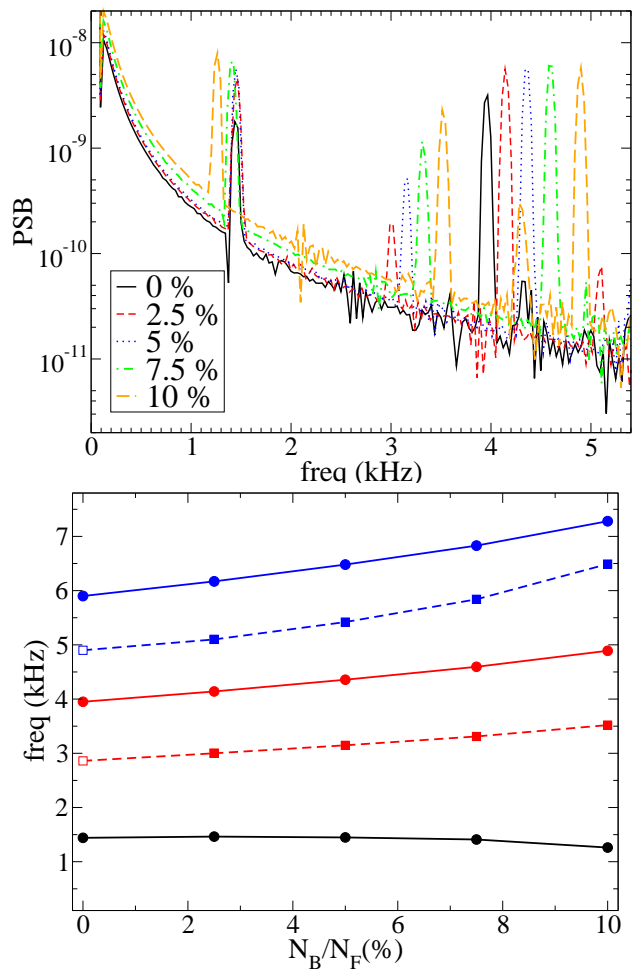

  \begin{center}
  \includegraphics[height=.27\textheight]{qnm.eps} \\
  \includegraphics[height=.27\textheight]{freq_Nb.eps}
  \caption{{\it Normal modes of the fermion-boson stars}
    (Top) Fourier spectrum of the central value of the fluid
    density $\rho_0$ for several stable configuration with
    $N_B/N_F=\{0,2.5,5,7.5,10\}\%$ and $M_T=1.4$.
    (Bottom) Frequencies corresponding to the first, second and third
    modes of the isolated neutron star, as a function of the 
    boson fraction. Notice the appearance of new oscillation modes,
    not present for an isolated neutron star. See the text for more details.}
  \label{fig:qnm}
  \end{center}
\end{figure}

We will restrict our analysis to a fermion-boson star with 
total mass $M_T=1.4$, and parameterize different mixed stars by
increasing the amount of bosons relative to fermions,
corresponding to the fractions $N_B/N_F=\{0,2.5,5,7.5,10\}\%$. The
details of the parameters of the stars are summarized in
Table~\ref{tab:stars}. We have evolved for long times $t \approx
6000$ in order to get at least $50$ oscillations of the central 
density, which will produce a clear spectrum with sharp peaks in the
frequency domain. The Fourier transform of this quantity is shown in
the top panel of Fig.~\ref{fig:qnm}. As an additional check of our
code, we can compare the known frequencies of the fundamental mode 
and its overtones for a (fermion-only) neutron star (as computed
either by using perturbation theory or numerical evolutions,
see for instance \cite{Font:2001ew})
with the ones obtained from our simulation for the purely fermionic
case (corresponding to the circles on the left in the bottom panel of
Fig.~\ref{fig:qnm}). The difference is always smaller than $1\%$,
confirming the accuracy and correctness of our results.

We now turn our attention to the boson-fermion case. The fundamental
mode, which is usually a function of the mean density of the star,
remains roughly constant except for the largest boson fraction, for
which it shifts towards smaller frequencies. The overtones, at higher
frequencies, display more interesting features with the presence of
new quasi-normal modes. The original neutron star overtones, displayed
with circles in Fig.~\ref{fig:qnm}, are the dominant ones for small
number of bosons. The power of these new oscillation modes increases
with the boson fraction, suggesting that their origin is the
gravitational coupling with the scalar field. The frequency of the
overtones has a significant drift towards higher values as the
fraction of bosons increases.

The main features of this spectrum can be qualitatively explained
in a very simple way. The new quasi-normal modes, which were not
present for isolated fermionic stars, corresponds to the quasi-normal
modes of the boson star. The oscillations in the bosonic part propagate
to the fermions through gravity. As the fraction of bosons
increases, so does the relative importance of the scalar field with respect
to the fluid density, producing the observed growth in the amplitude of
these modes. Consequently, the spectrum can be mainly understood as
a superposition of the quasi-normal modes of the boson and the neutron star.
The drift in the frequencies is an effect of the change in radius and
mean-density of the star as the fraction of boson changes.


\section{Concluding remarks}
\label{sec:concluding_remarks}

We have studied in some detail the numerical evolution of equilibrium
configurations of mixed boson-fermion stars. Our results confirm
the existence of stable and unstable branches of equilibrium
configurations. We also defined a stability criterion based on the
variation of the number of bosonic and fermionic particles, for a given
fixed value of the total mass, as a function of the central values of the
scalar field amplitude and the fluid density. This criterion states
that the
equilibrium configurations located on the left of the maximum (minimum)
number of bosons (fermions) are stable, whereas the configurations located
on the right are unstable. We were able to determine the curve that separates the
stable branch from the unstable one, in the plane formed by the
central values of the scalar field $\phi_c$ and fluid density $\rho_c$.
We also verified that the correct solutions are
obtained in the limiting cases of an isolated boson or fermionic star,
by comparing with the results of previous studies. 

In order to assess the stability criterion, we performed the numerical
evolution of the fully non-linear equations of motion for two types of
solutions. For the stable configuration, the central values of the
scalar field and the fermionic density remain constant in time during
the numerical evolution, while the unstable star migrates to a stable
configuration by ejecting out some of the initial mass.

We also studied the structure of the normal modes and overtones of
these mixed stars by performing long term numerical evolutions for
configurations with a fixed total mass but with different boson to
fermion ratios. As expected, new oscillation modes appear in the
frequency spectrum of the stars, when compared to the fermion-only case;
the appearance of the new overtones is justified because of the
gravitational coupling of the fermionic perfect fluid with the scalar
field, which has its own oscillation modes.

As we mentioned before, an accurate classification of the properties
of boson-fermion stars is necessary in order to investigate the possible
existence of bosons trapped inside, for instance, in neutron
stars. One possible indication of such a phenomenon would be the
shift in frequency and the presence of new modes in the vibration
spectrum of the stars. 

Another case of astrophysical interest is the possible
existence of bosonic dark matter in galactic halos, an idea that has
drawn some attention in the specialized literature in recent
years~\cite{UrenaLopez:2010zz,Bernal:2009zy,Matos:2008ag,Matos:2004rs,
  Matos:2000ss,Matos:2000ng,Matos:1999et,PhysRevLett.85.1158,PhysRevD.53.2236,PhysRevD.50.3650,PhysRevD.62.103517,PhysRevD.64.123528}. This  
type of models refers to the other extreme case, in   
which the star is dominated by the bosonic component. In the context
of galaxies, these mixed fermion-boson stars could model the galaxy
halo with a boson star, and the gas with a fermionic component.
Our present results indicate that a boson-dominated galaxy halo must
keep its stability features after the inclusion of fermions; however,
more work is needed to determine the properties of the equilibrium configuration
that may be detected through astrophysical
observations. This is work in progress that we expect to report
elsewhere.

%
%

\begin{acknowledgments}
We are grateful to Juan Barranco, Cecilia Chirenti, Luis Lehner and
Steve Liebling for useful comments and discussions. SV-A acknowledges
support from CONACyT, M\'exico, and the kind hospitality of the
Canadian Institute for Theoretical Astrophysics (CITA, Canada) for a
research stay where part of this work was done. This work was
partially supported by PIFI, PROMEP, DAIP-UG, CONACyT M\'exico under
grant 167335, and the Instituto Avanzado de Cosmologia (IAC)
collaboration. DA acknowledges support from the DFG grant SFB/Transregio~7.

\end{acknowledgments}

%
%

\appendix

\section{The evolution system of equations}
\label{app:z3}

We consider the Z3 formulation of the Einstein equations in spherical
symmetry~\cite{Bernal:2009zy} as the evolution system for the
space-time geometry. The system is regularized at the origin using the
following transformation of the momentum constraint: 
\begin{eqnarray}
\tilde{Z_{r}} &=& Z_{r} +
\frac{1}{4r}\left(1-\frac{g_{rr}}{g_{\theta\theta}}\right),\nonumber
\end{eqnarray}
which ensures the cross-cancellation of the factors $1/r$ in the fluxes, and
$1/r^{2}$ in the sources. The sources still have terms like $1/r$
times other variables that contain radial derivatives of the metric
coefficients. However, these terms do not create problems at $r
\rightarrow 0$, as the radial derivatives of any differentiable
function must vanish at the origin. Thus, the equations of motion read
\begin{widetext}
\begin{subequations}
\label{eq:forz3}
\begin{eqnarray}
  \partial_{t}A_{r} &=& -\partial_{r}[\alpha \, trK] \,, \label{Ar} \\
  \partial_{t}D_{rr}{}^{r} &=& -\partial_{r}[\alpha \, K_{r}{}^{r}] \,, \label{drr} \\
  \partial_{t}D_{r\theta}{}^{\theta} &=& -\partial_{r}[\alpha \,
  K_{\theta}{}^{\theta}] \,, \label{dthetatheta} \\
  \partial_{t}Z_{r} &=& -\partial_{r}[2 \, \alpha \, K_{\theta}{}^{\theta}] +
  2\alpha\Big\{(K_{r}{}^{r}-K_{\theta}{}^{\theta})\Big(D_{r\theta}{}^{\theta}
+ \frac{1}{r} \Big) - K_{r}{}^{r} \Big[ Z_{r} + \frac{1}{4r} \Big( 1 -
\frac{g_{rr}}{g_{\theta\theta}} \Big) \Big]  \nonumber\\
 & & + A_{r}K_{\theta}{}^{\theta}+\frac{1}{4r}\frac{g_{rr}}{g_{\theta\theta}}
  (K_{\theta}{}^{\theta}-K_{r}{}^{r}) -4\pi S_{r}\Big\} \, , \label{zr} \\
\partial_{t}K_{r}{}^{r} &=& -\partial_{r}\Big[\alpha g^{rr}\Big(A_r+
\frac{2}{3}D_{r\theta}{}^{\theta}-\frac{4}{3}Z_{r}\Big)\Big]+
\alpha\Big\{(K_{r}{}^{r})^{2}+\frac{2}{3}K_{\theta}{}^{\theta}(K_{r}{}^{r}-K_{\theta}{}^{\theta}) \nonumber\\
 & & - g^{rr}D_{rr}{}^{r}A_{r} +\frac{1}{3r}[g^{rr}(D_{rr}{}^{r}-A_{r}-4Z_{r})+g^{\theta\theta}
 (D_{r\theta}{}^{\theta}-A_{r})] \nonumber\\
 & & + \frac{2}{3}g^{rr}\Big[Z_{r}+\frac{1}{4r}\Big(1-\frac{g_{rr}}
 {g_{\theta\theta}}\Big)\Big](2D_{rr}{}^{r}-2D_{r\theta}{}^{\theta}-A_{r}) \nonumber\\ 
 & & - \frac{2}{3}g^{rr}\Big(D_{r\theta}{}^{\theta}+
\frac{1}{r}\Big)(D_{rr}{}^{r}-A_{r})+
8\pi\Big(\frac{\tau}{6}-\frac{S_{r}{}^{r}}{2}+S_{\theta}{}^{\theta}\Big)\Big\}
\, ,\label{krr}\\
\partial_{t}K_{\theta}{}^{\theta} &=& -\partial_{r}\Big[\alpha
g^{rr}\Big(-\frac{1}{3}D_{r\theta}{}^{\theta}+\frac{2}{3}Z_{r}\Big)\Big]+
\alpha\Big\{\frac{1}{3}K_{\theta}{}^{\theta}(-K_{r}{}^{r}+4K_{\theta}{}^{\theta})\nonumber\\
 & & +  \frac{1}{6r}[g^{rr}(A_{r}-2D_{rr}{}^{r}-4Z_{r})+g^{\theta\theta}(A_{r}-2D_{r\theta}{}^{\theta})] \nonumber\\
 & & - \frac{2}{3}g^{rr}\Big[Z_{r}+\frac{1}{4r}\Big(1-
\frac{g_{rr}}{g_{\theta\theta}}\Big)\Big](D_{rr}{}^{r}-D_{r\theta}{}^{\theta}-2A_{r})\nonumber\\
 & & + \frac{1}{3}g^{rr} \Big(D_{r\theta}{}^{\theta}+\frac{1}{r} \Big)(D_{rr}{}^{r}-4A_{r})+ 
8\pi\Big(\frac{\tau}{6}-\frac{S_{r}{}^{r}}{2}+S_{\theta}{}^{\theta}\Big)\Big\}
\, ,\label{kthetatheta}
\end{eqnarray}
\end{subequations}
\end{widetext}
where $Z_{r}$ is the vector associated with the Z3 formulation, and
$trK=K^{r}{}_{r}+2K^{\theta}{}_{\theta}$ is the trace of the extrinsic
curvature. In Sec.~\ref{sec:formalism} we defined the  matter terms of
the fermionic fluid \{$D$, $U$, $\tilde{S}_r$, $\tilde{S}_r{}^r$,
$\tilde{S}_{\theta}{}^{\theta}$\}, and the auxiliary variables
\{$A_r$, $D_{rr}{}^r$, $D_{r\theta}{}^{\theta}$\} which we introduced
in order to reduce the full system in Eqs.~\ref{eq:ekghd},
\ref{eq:lapse}, and~\ref{eq:forz3}, to first order in space and time.

The total matter terms are given by
\begin{subequations}
\begin{eqnarray}
  \tau &=& \frac{1}{2}(g^{rr}\phi^{*}_{t}\phi_{t}+g^{rr}\phi^{*}_{r}
  \phi_{r}+V(\phi))+U\, , \\
  S_{r} &=&-\frac{1}{2}[\sqrt{g^{rr}}\phi^{*}_{t}\phi_{r}+\sqrt{g^{rr}}
  \phi_{t}\phi^{*}_{r}] + \tilde{S}_{r}\, , \\
  S_{r}{}^{r} &=&\frac{1}{2}[g^{rr}\phi^{*}_{t}\phi_{t}+g^{rr}
  \phi^{*}_{r}\phi_{r}-V(\phi)] + \tilde{S}_{r}{}^{r} \,
  , \\
  S_{\theta}{}^{\theta} &=& \frac{1}{2}[g^{rr}\phi^{*}_{t}\phi_{t}-g^{rr}
  \phi^{*}_{r}\phi_{r}-V(\phi)]+\tilde{S}_{\theta}{}^{\theta}\,.
\end{eqnarray}
\end{subequations}
The total mass of the mixed stars is calculated from the 
ADM mass defined as
\begin{equation}\nonumber
M_{ADM} = \frac{1}{16\pi} \lim_{r \to \infty} \int g^{pq}[\partial_{q}g_{pk} -
\partial_{k}g_{pq}] N^{k} dS,
\end{equation}\nonumber
where $N^{r}=\sqrt{g^{rr}}\delta_{r}{}^{r}$ is the unit outward normal to the sphere.
In our coordinates, it can be translated into
\begin{equation}
M_{ADM} = - r^{2} \sqrt{g^{rr}} Dg_{r\theta}{}^{\theta}.
\end{equation}
For stable stars, we also use the Tolman
mass defined as
\begin{eqnarray}
M_{T} &=& \int (T_{0}{}^{0}-T_{i}{}^{i})\sqrt{-g}dx^{3} \, , \\
 &=& 4\pi \int r^2\alpha\sqrt{g_{rr}}g_{\theta\theta}(\tau + S_{r}{}^{r} 
+ 2S _{\theta}{}^{\theta}) dr \,.\nonumber
\end{eqnarray}
On the other hand, the number of fermionic particles associated to the
mass of the fermionic fluid is given by
\begin{equation}
N_F= 4\pi \int r^2\sqrt{g_{rr}}g_{\theta\theta} (\rho W) dr \,.
\end{equation}
The number of bosonic particles can be associated to the Noether
charge~\cite{Ruffini:1969qy} of the scalar field, which can be computed as 
\begin{equation}
N_B = 4 \pi \int \frac{r^2}{2 i \alpha} \sqrt{g_{rr}}g_{\theta\theta}
(\phi^{*} \partial_{t} \phi - \phi \partial_{t} \phi^{*}) dr \, .
\end{equation}

The Hamiltonian constraint takes the form
\begin{eqnarray}
H &=& \frac{2}{g_{rr}} \Big\{ -2 \partial_i D_{r \theta}{}^{\theta} - 3 D_{r \theta}{}^{\theta}
\Big(D_{r \theta}{}^{\theta} + \frac{2}{r} \Big) \nonumber\\
& & + g_{rr} K_{\theta}{}^{\theta} (K_{\theta}{}^{\theta} + 2 K_{r}{}^{r}) 
- \frac{(1 - g_{rr} g^{\theta\theta})}{r^2} \nonumber\\
& & + 2 D_{rr}{}^{r} \Big(\frac{1}{r} 
+ D_{r\theta}{}^{\theta} \Big) - 8 \pi g_{rr} \tau \Big\}. \label{eq::ham}
\end{eqnarray}

\section{The transformation from conserved to primitive quantities}
\label{app:conservadas}

From the definition of the conserved quantities
\begin{equation}\label{eq:conserved}
   D = \rho_{0} W\,, \quad  U = h W^{2} - P \,, \quad 
   \tilde{S}_{r} = h W^2 v_{r} \, ,
\end{equation}
one obtains the primitives $\{\rho\, , P\, , v_r\,, \epsilon\}$ after
each time integration of the equations of motion. This is not trivial,
mainly because the enthalpy $h=\rho(1+\epsilon)+ P$, and the
Lorentz factor $W=1/\sqrt{1-v^rv_r}$, are defined as functions of the
primitives. 

We are adopting a recovery procedure which consists in the following steps:
\begin{enumerate}
\item From the first thermodynamics law for adiabatic processes, it
  follows that
\begin{equation}\label{presion}
P = (\Gamma -1) \rho\epsilon \, .
\end{equation}
Substituting the definition of the entalphy in the equation of state above, 
we write the pressure as a function of the conserved quantities and
the unknown variable $x = h W^2$. 

\item Using the previous step, the definition of $U$ becomes:
\begin{eqnarray}\label{eq:forx}
   U &=& hW^2 - P\nonumber\\ 
   &=& hW^2 -\frac{(\Gamma -1)}{\Gamma}(h-\rho)\nonumber\\
   &=& hW^2\Big(1-\frac{\Gamma-1}{\Gamma}\Big)+
   \frac{\Gamma-1}{\Gamma}\rho \, ,
\end{eqnarray}
where $\Gamma$ is the adiabatic index corresponding to an ideal gas.

\item Then, the function
\begin{eqnarray}
f(x) &=& \left(1 - \frac{\Gamma - 1}{W^2 \Gamma} \right) x +
\frac{D (\Gamma - 1)}{W \Gamma} - U,
\end{eqnarray}
must vanish for the physical solutions. The roots of the function
$f(x)=0$ can be found numerically by means of an iterative
Newton-Raphson solver, so that the solution at the $n+1$-iteration can
be computed as
\begin{equation} 
x_{n+1} = x_n - \frac{f(x_n)}{f'(x_n)},
\end{equation} 
where $f'(x_n)$ is the derivative of the function $f(x_n)$. The
initial guess for the unknown $x$ is given in the previous time step.

\item After each step of the Newton-Raphson solver, we update the
  values of the fluid primitives as
\begin{equation}\label{eq:primitive}
   \rho = D/ W\,, \quad P = x - U \,, \quad 
   v_{r} = \tilde{S}_{r}/x \, , \\ 
\end{equation}
where $W^2=x^2/(x^2-\tilde{S}^r\tilde{S}_r)$.

\item Iterate steps 3 and 4 until the difference between two
  successive values of $x$ falls below a given threshold value of the
  order of $10^{-10}$.
\end{enumerate}

%
\bibliography{paper}

\end{document}